# Smart Watch Supported System for Health Care Monitoring

Anshuman Mishra, Richards Joe Stanislaus, *Member, IEEE*
*Vellore Institute of Technology,*
*India*

*Abstract*— This work presents a smart watch attached to patient at remote location, which would help in navigation of wheel chair and monitor the vitals of patients and relay it through IoT. This wearable smart watch is equipped with sensors to measure the health parameters: heartbeat, blood pressure, body temperature and step count. An esp8266 Wi-Fi module uploads the health parameters into the thingspeak cloud platform with time stamp. This smart watch is equipped with joystick for cruise and navigation control of the motor driver enabled wheel chair. Additionally, an ultrasonic sensor mounted in front of wheel chair continuously scans for any obstacles ahead and stops the motion of the wheel chair on detection of an obstacle. The system's primary controller is an Arduino UNO microcontroller, which interfaces the input and output modules.

## I. INTRODUCTION

The cost of health care has increased in the modern era, and constant supervision of aged and dependent family members in their homes is essential in today's hectic society. At the moment, wearable smart watches are considered a medical technology breakthrough. Recently, expensive smart watches have been outfitted with the most basic diagnostic, therapeutic, and patient monitoring features. Recently, it has been proposed and put into practise to have the ability to monitor physical activity, sleep patterns, pulse rate, heartbeats, oxygen saturation, and blood pressure. The smart watches can also be used to set alarms for regular activities like exercising and taking prescriptions. Few designs incorporate the monitoring of physical activity, including location and daily step count. Wearable technology will undoubtedly alter our lives, especially for early adopters. Many businesses have committed money in creating wearable with distinctive features in order to efficiently access buyer markets. Given that the arm moves more frequently than any other body component and can therefore track all movements of the body, the wrist is the optimal spot to show all essential data. The wrist has veins and arteries that can be used to gauge blood pressure and pulse, and the skin is thin enough to accommodate sensors. An attachment that can be worn as a smart watch or covertly put into a patient's clothing is a small electronic gadget. [3]. The majority of wearable technology uses contemporary biosensors and remote information communication, which allow for data access and transmission in all areas of human endeavour. Given the utilisation of smaller biosensors equipped for remote connection, these devices are intended to be inconspicuous, innovative observation innovations for continuous and independent transfer of physiological information [4]. As wearable devices expand in the clinical setting, they may equip professionals with the knowledge they need to enhance patient care, clinical workflow, and remote patient treatment, collect more detailed health data, and offer patients more thorough medical care [5]. The development of numerous crucial components, including normalisation, digitalization, systems administration, reconciliation, and scaling down, was recognised by Zhang's study team as being necessary for wearable technology to perform [6]. The clever is a completely new design for the rapidly expanding industry of wearable technology. A smart's streamlined figuring innovation and structure factor plan enable continuous wear without impeding the client's daily activities. PDAs are now commonplace and might be categorised as wearables, but they are normally kept in a bag or pocket. Unlike PDAs, smart watches can be worn without interfering with daily activities and can also be a convenient addition to a current cell phone [8]. The medical care business has endless potential for smart watches with biosensors, and they may be able to give patients and their suppliers vital information about medical services.

Smart watches are a type of wearable computer that mimics a watch. The touchscreen interface of modern smart watches allows for ease of use on a daily basis, and an app on a smartphone is used to run the device and collect telemetry data for uses like long-term bio monitoring. Early iterations of smart watches could perform basic functions including math calculations, digital timekeeping, game play, and translation. Smart watches have evolved features more equivalent to WiFi/Bluetooth connectivity, a mobile operating system, including mobile apps, and smartphones since its introduction in 2015. Some smart watches can play digital video and audio files over Bluetooth and FM radio headsets, making them portable media players. Some variations—known as watch phones—incorporate mobile cellular functionality, such as the ability to make phone calls. The suggested smart watch system includes three features: obstacle detection, wheel chair control with a joystick capability, and IOT technology for health monitoring.

THINGSPEK CLOUD:

The Internet of Things platform and electronics are this project's main research areas. The fact that Arduino produces the output needed by the user after receiving input from the sensors makes it a crucial part of this project.

The Internet of Things is the technology that is growing the quickest. Everything and everything is about to use the Internet of Things. Internet of Things adoption will improve medical electronics. As a result, the Internet of Things platform Thing speaks is used. ThingSpeak is among the best resources for IoT projects. Among other things, we may use this platform to monitor our data and control the internet system through channels. In this pandemic situation, doctors are unable to continuously check a patient's body temperature and heart rate. Therefore, we made the decision to conduct this project in order to make the task of hospital management easier. A heartbeat sensor and an ESP8266 WiFi modem are used in this project. A linked Arduino Uno receives data from sensors that measure temperature, blood pressure, and steps and sends it to the Thingspeak cloud.

This low-tech but highly effective apparatus allows for continuous monitoring of the health of a seriously ill patient. By referring to all open-source platforms, we suggested a tool called Thing Speak that is easier to use, safe, and offers a variety of options. Data that maintains patient records updated and is available whenever needed can be saved and retrieved via Thing Speak. The heartbeat and pulse can be monitored independently, and we can export or import solo or combined sets of data as well as compare data using visualisations.

## WHEEL CHAIR CONTROL

The wheelchair is very beneficial for elderly people and those with physical disabilities. Robotics and intelligent system technology can be used to produce the propelled wheelchair. The wheelchair's joystick control system makes it very easy to operate. Using the joystick option, a physically challenged person can control the wheelchair's movement. In this project, we used a microcontroller to manage and monitor the system.

The command is carried out using the Arduino controller by the microcontroller after being transmitted to it through a joystick. The L293D motor driving IC receives a digital signal command from the controller once it has been executed, and this IC in turn controls the motion of the two dc motors. As a result, the dc motor spins in response to the joystick's order.

### OBSTACLE DETECTION SYSTEM:

Industrial applications for ultrasonic sensors are distinguished by their exceptional adaptability and dependability. Ultrasonic sensors can be used to handle even the most challenging jobs, including level measurement or object detection with millimetre precision, because its measurement method is virtually always precise. In many commonplace items, infrared sensors are used. They are coveted for their low power requirements, portability, and straightforward circuitry.

Sound waves from the ultrasonic sensor are picked up by sound waves that reverberate off of objects. The energy of the ultrasonic waves is diffusely reflected over a wide solid angle, which can be as high as 180 degrees, when they strike with an object. The incident energy is thus partially reflected back to the transducer. When the object is very close to the sensor, the sound waves quickly return; however, when the object is farther away, the sound waves return more slowly. The signal's return time will be too long or it will be too weak for the receiver to pick up if an object is too far away from the sensor.

Based on how long it takes for the sound to return from the object in front of the sensor, it may determine how far away anything is.

$$L = \frac{vt \cos \theta}{2}$$

Then, using the relationship where "t" denotes the amount of time the wave takes to return to the sensor and "θ" denotes the angle between the path taken and the horizontal, it is possible to determine the distance to the object (L) using the speed of ultrasonic waves (v) in the medium, as shown in the figure. Instruments based on the Doppler shift are employed while the item is moving. The ultrasonic sensor can measure distances in inches and centimetres. It can measure distances between 0 and 2.5 metres with an accuracy of 3 cm.

## III IMPLEMENTATION

*3.1. Block diagram of Smart Watch*

We developed a wearable, Arduino-powered smart watch. ESP8266 WiFi module with heart rate monitoring. DC motors, ultrasound, step count, blood pressure, and temperature sensors, as well as sensors for L293D motor drivers. The Arduino UNO will continuously read data from health sensors and update it with time and date in the Thingspeak cloud through the esp8266 WI-FI module. When a user steers the wheelchair with a joystick, the Arduino microcontroller manages the wheelchair's directions by managing the DC motors. If the wheelchair runs into any obstructions, it will immediately stop moving thanks to an ultrasonic sensor. A microcontroller loaded with embedded C language does this work.

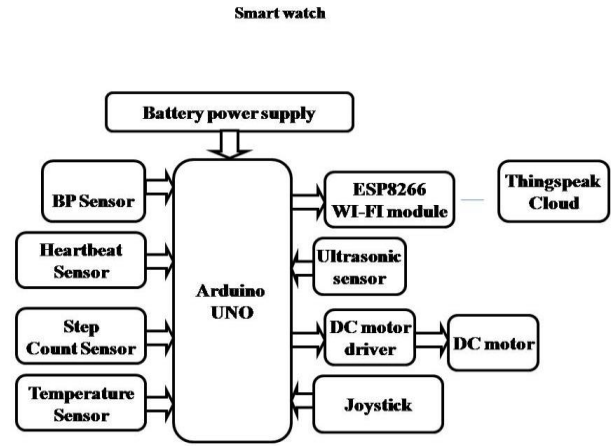

Fig. 1. Block diagram of proposed system

## IV. METHODOLOGY and RESULT

The following discussion gives a brief overview to the various modules utilized in this project:

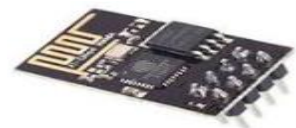

Fig 2: Esp8266 Wi-Fi module

*4.2. ESP8266 WI-FI module with Thingspeak:*

Data from sensors, websites, and devices may be transmitted to the cloud via Thing Speak and stored in either a public or private channel. As soon as information is entered into a Thing Speak channel, it may be visualised, analysed, utilised to create new information, or used to interact with other devices, online communities, and web services.

There are three IoT architectural layers that make up the project's implementation.

three layers: the physical, network, and application layers.

Physical layer – I can sense some of the setup. The setup and use of sensors to measure physical parameters are included in the physical Layer. In this case, every health sensor obtains the necessary data straight from the patient's body. Sensors collect data on physical quantities, which they subsequently transmit to transducers for transduction, or, more precisely, for converting the signals into electrical inputs. They are then converted into electrical form and fed into two different input pins on the Arduino. The output from the analysis will be delivered to the ESP2866 Wi-Fi module using the code that has been loaded into the Arduino.

Network Layer – Data transfer from a device to Thing Speak. The network layer describes the interaction and linkage between the IoT platform and the active device. We use a Wi-Fi module to connect to and communicate with the device, in this case Thing Speak, for data analysis. The cloud platform and the module will be able to communicate by using the necessary module commands (ESP2866 instructions). To begin the data transfer, repeat the process, put an end to the transfer, and deliver the data, use the aforementioned commands.

Application Layer: Thing Speak is used for continuous monitoring because it is a cloud platform that can be accessed from anywhere at any time and is also user-friendly for monitoring health parameters in urgent scenarios. The data can be shown graphically and in a variety of different (user defined) ways.

*4.3. Arduino Uno:*

Arduino Uno: The Arduino Uno microcontroller board is built around the ATmega328. The device has 20 digital input/output pins, six analogue inputs, six PWM outputs, a 16 MHz resonator, an in-circuit system programming (ICSP) header, a power jack, and a USB port.

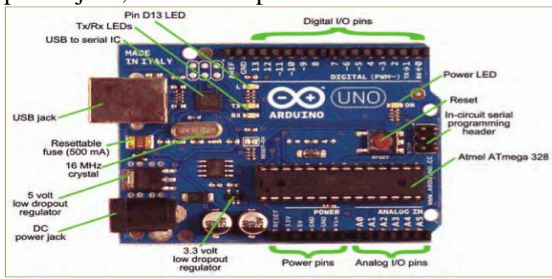

Fig 3 Arduino Uno

4.4. Heartbeat Sensor and *Temperature sensor:*

A sensor system called the MAX30100 features integrated pulse oximetry and heart rate monitoring. A sensor takes in a physical quantity and converts it into a signal that may be understood by a viewer or instrument. The output of the LM 35 temperature sensor is expressed in millivolts (mv).

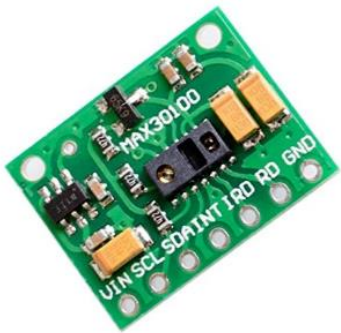 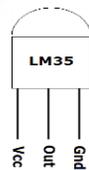

Fig4a Heartbeat sensor     4.b LM35 Temperature Sensor

*4.5. DIGITAL BP SENSOR:*

The totality of colours that can be depicted in the mentioned medium is referred to as the "colour space." For our purposes, grayscale, RGB (red/green/blue), and CMYK (cyan/magenta/yellow/black) are the three primary colour spaces. While CMYK is utilised for printing, RGB is typically employed with on-screen graphics.

The colour spaces RGB or CMYK should be used to create all colour figures. Images in grayscale colour space should be supplied. Grayscale or bitmap colorspace are both acceptable options for line art. Be aware that "bitmap file format" and "bitmap colorspace" are not the same thing. When bitmap colour space is chosen, the suggested file formats are .TIF/.TIFF/.PNG.

The force that flowing blood exerts on blood vessel walls is known as blood pressure (BP).

*4.6. STEP COUNT SENSOR with Ultrasonic Sensor:*

The sensor is made of a silicon wafer with a micro-machined structure on it. Polysilicon springs that are used to support the structure allow for smooth deflection in any direction when the X, Y, and/or Z axes are accelerated. An instrument called an inaudible sensing element uses inaudible sound waves to gauge the distance between a target object and itself before converting the reflected sound into an electrical signal.

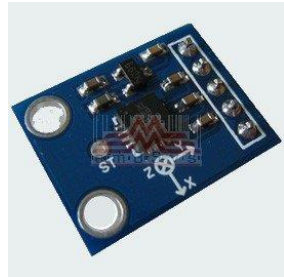 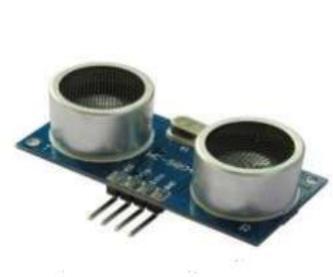

Fig. 5a MEMS sensor          Fig. 5b   Ultrasonic sensor

4.7. Joystick:

An analogue joystick, also referred to as a control stick, The joystick or thumb stick on a controller is a two-dimensional input device. An analogue joystick is analogous to two potentiometers, one for horizontal (X-axis) movement and the other for vertical (Y-axis) movement. Along with the joystick, there is a choice switch. It can be quite helpful for operating robots, RC cars, or older video games.

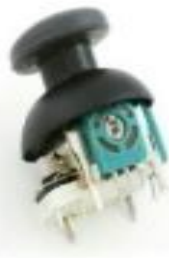 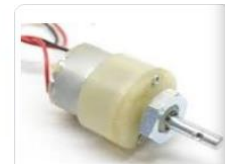

Fig 6a. Joystick           Fig 6b DC motor

4.10: DC motor with L293Driver  and L293D IC:

Using two DC motors, the wheelchair may be moved in many directions, including right, left, reverse, and forward. A microprocessor controls these motors.  Two integrated H-bridge driver circuits are present in L293D. Two DC motors can be run concurrently in both forward and reverse motion in the conventional mode of operation. Enable pins 1 and 9 (which relate to the two motors) must be high in order for motors to begin operating.

On integrating the system, we have demonstrated a successful health-monitoring system with the capabilities of wheel-chair control which care be remotely monitored. Figures 7 through 9 establish the remotely obtained parameters and the system connections.

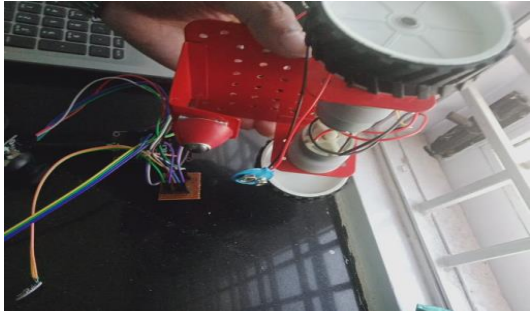

Fig. 7. a Interfacing sensors with motor system

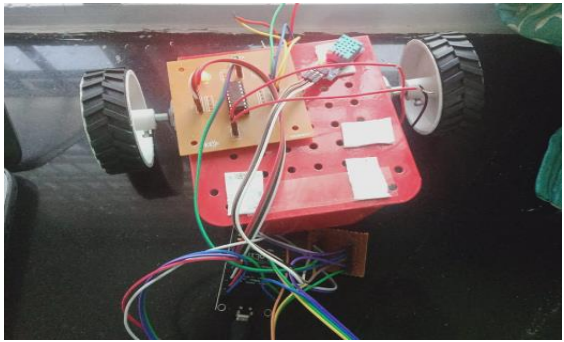

Fig 7b. Health-care sensors interfaced with arduino

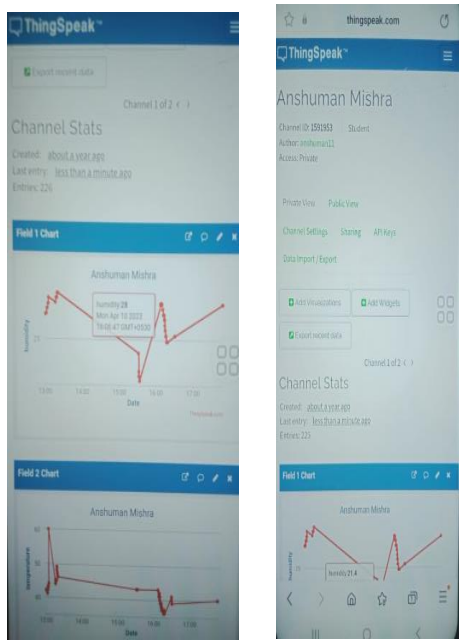

Fig. 8 Thingspeak output displaying the relayed health status of the patient.

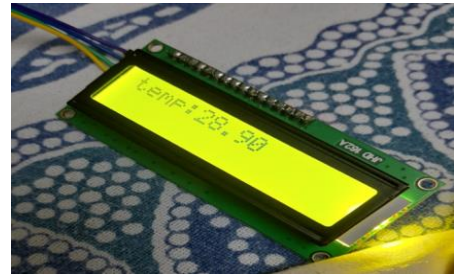

Fig. 9. Display module for displaying the health parameters.

## V. CONCLUSION

We have developed a smart watch system that can control a wheelchair, track health data, and clear obstructions. Health data can be tracked by this wearable smart watch in a remote location that is accessible from anywhere in the world. This wearable smart watch can be operated with the hands by those who are unable to walk, and it can recognise obstacles in its route and stop the wheelchair on its own.

## VII ACKNOWLEDGEMENT

We would like to thank all of the writers of the research articles that served as our paper's references. Significant knowledge was gathered, and it will be helpful in the future.